\documentclass[journal]{IEEEtran}

\usepackage{graphicx}
\usepackage{cite}
\usepackage{pbox}
\usepackage{amsmath}
\usepackage{multirow}
\usepackage{algorithm}
\usepackage[noend]{algpseudocode}
\usepackage{amssymb}
\usepackage{amsthm}
\usepackage{bm}
\usepackage{caption}
\usepackage{subcaption}
\usepackage{verbatim}
\usepackage{soul,color}
\usepackage{array}

\ifCLASSINFOpdf
\else
\fi
\makeatletter
\def\BState{\State\hskip-\ALG@thistlm}
\makeatother

\newcolumntype{P}[1]{>{\centering\arraybackslash}p{#1}}
\newcolumntype{M}[1]{>{\centering\arraybackslash}m{#1}}

\begin{document}

\title{Profit-aware Online Vehicle-to-Grid Decentralised Scheduling under Multiple Charging Stations}

\author{Abbas~Mehrabi,~\IEEEmembership{Student Member,~IEEE,}
        Aresh~Dadlani,
        Seungpil~Moon,
        and~Kiseon~Kim,~\IEEEmembership{Senior~Member,~IEEE} 
\thanks{A. Mehrabi, A. Dadlani, and K. Kim are with the School of Electrical Engineering and Computer Science, Gwangju Institute of Science and Technology (GIST), Gwangju 61005, South Korea (e-mail: \{mehrabi,dadlani,kskim\}@gist.ac.kr).}
\thanks{S. Moon is with Korea Electric Power Research Institute (KEPRI), Daejeon 305-760, South Korea (e-mail: spmoon@kepri.re.kr).}
}

%

\maketitle

\begin{abstract}
Fluctuations in electricity tariffs induced by the
sporadic nature of demand loads on power grids has initiated
immense efforts to find optimal scheduling solutions for charging
and discharging plug-in electric vehicles (PEVs) subject to
different objective sets. In this paper, we consider vehicle-togrid (V2G) scheduling at a geographically large scale in which
PEVs have the flexibility of charging/discharging at multiple
smart stations coordinated by individual aggregators. We first
formulate the objective of maximizing the overall profit of both,
demand and supply entities, by defining a weighting parameter.
We then propose an online decentralized greedy algorithm for
the formulated mixed integer non-linear programming (MINLP)
problem, which incorporates efficient heuristics to practically
guide each incoming vehicle to the most appropriate charging
station (CS). The better performance of the presented algorithm
compared to an alternative allocation strategy is demonstrated
through simulations in terms of the overall achievable profit
and flatness of the final electricity load. Moreover, the results of
simulations reveal the existence of optimal number of deployed
stations at which the overall profit can be maximized.
\end{abstract}


%
\IEEEpeerreviewmaketitle

\section{Introduction}
\label{sec:introduction}

\IEEEPARstart{G}{rowing}  awareness of energy and environmental crises
has catalyzed the evolutionary shift towards electrifying
personal transportation. In recent years, investments on electric
vehicles (EVs) as eco-friendly and cost-efficient substitutes for
conventional fuel-propelled automobiles that exhaust natural
resources have been over-whelming. Classified broadly based
on their mode of propulsion, plug-in electric vehicles (PEVs)
which are purely battery electric vehicles have benefits far
more broad than conventional vehicles [1]. These benefits
however, are accompanied by various new challenges as more
EVs are integrated into the power grid. Prime concerns include
power distribution instability and transmission congestion due
to the unmanaged charging and/or discharging of EVs at
grid-connected electric vehicle supply equipments, commonly
known as charging stations (CSs). 

From the research prospective, optimal charging and discharging of EVs have been widely scrutinized due to their
significant impact on the load regulation of vehicle-to-grid
(V2G) systems [2], [3], [5], [6]. The bidirectional flow of
power between EVs and power grid facilitates load flattening by shifting the demands of charging EVs from peak load
hours to off-peak periods [7]. With regard to power costs at
different time intervals and uncertainty in EV arrival times,
smart scheduling techniques are required to not only satisfy
profit expectations, but also prevent the grid from crashing
through power load flattening [3], [9]. The geographical scale
over which existing works on V2G scheduling are studied can
be classified in terms of the number of aggregators involved.
Non-preemptive allocation of EVs for charging/discharging
operations occur either at a single CS managed by a single
aggregator (SCS-SA) [3], [9], [10], where the scheduler has
global information on the energy requirement and departure
time of each EV, or at multiple CSs coordinated by an
aggregator (MCS-SA) [3], where the scheduling optimization
problem is locally solved in each group.

The V2G scheduling of EVs, comprising of multiple number
of spatially-distributed CSs each managed by individual aggregators has however, not yet been investigated under which
EVs seeking service experience a higher degree of flexibility in selecting the station that yields the most achievable
profit. Under this scenario, the overall profit can be affected
by variations arising in the system parameters such as CS
cardinality and maximum vehicle capacity at each CS. Hence,
selection of optimal system parameters can essentially alleviate
the net auxiliary and establishment costs incurred by the CSs.
Moreover, the authors of [3] and [9] merely focus on the
gross profit of EV owners without accounting for the profit
of CSs in their objective functions. On the other side, the
EVs scheduling problem in [22] considers the objective of
maximizing the overall obtainable profit of only aggregators.
From the V2G management system point of view, in order
to encourage the energy utility provider to establish CSs for
delivering the energy to EVs, it necessitates to share relatively
the obtainable profit between EVs and the CSs.

Inspired by the before-mentioned limitations and this later
motivation, the main contributions of this paper are highlighted
as follows:
\begin{itemize}
\item The problem of profit maximization considering realtime pricing for EV charging/discharging scheduling in
a large-scale V2G system composed of multiple stationsmultiple aggregators (MCS-MA) is formulated. A mixed
integer non-linear programming (MINLP) optimization
model is then proposed for the problem formulation
which, in contrast to the previous models, accounts for
adjustable profit between EV owners and CSs. 
\item To cope with the scheduling problem involving stochastic real-time EV arrivals, an online greedy algorithm
employing internal heuristics is proposed to guide each incoming EV to the most profitable station. The proposed
algorithm has low time and message complexities per
vehicle, which makes it applicable for large-scale V2G
deployments. 
\item Outperformance of the proposed algorithm in comparison
to an alternative allocation strategy in terms of overall
achievable profit as well as the flatness of final electricity
load are shown through simulation results. Furthermore,
optimal point for the number of deployed stations is also
attained by system parametric adjustments. This optimal
point can help V2G system designers to optimize their
investment budget.
\end{itemize}
The remainder of this paper is structured as follows. Related
works on EV scheduling in smart grids are briefly reviewed
in Section II. Section III introduces the system model and
notations, followed by scheduling problem formulation in
Section IV. The proposed online greedy algorithm and its
theoretical analysis are detailed in Section V. Simulation setup and results are provided in Section VI. Finally, Section VII
concludes the paper.

\section{Related Work}

Allocation strategies for EV charging/discharging with different objectives have been studied in many recent works
[2]- [5], [9]- [11]. The main challenge in devising real-time
allocation algorithms in V2G is the uncertainty of future
departure times and charging demands of EVs a priori. The
non-preemptive scheduling problem studied by He et al. [3]
accounts for the real-time pricing and degradation/fluctuation
costs of batteries in obtaining the minimum charging costs.
They consider the scenario of online EVs arrival to several
small and closely-located CSs managed by a single aggregator
and design a decentralized locally-optimal algorithm. In [11],
the authors provide a closed-form solution to determine the optimal charging power of a single EV under time-of-use (ToU)
pricing model and uncertain departure time. The authors of [5]
proposed an online algorithm with proven competitive ratio for
obtaining a sub-optimal solution with slightly higher cost as
compared to the offline optimal solution, while satisfying the
desired energy requirements imposed by the vehicles.

While many other set-ups and approaches have been applied
to envisage vehicle-to-grid interactions in the literature [12],
[13], scheduling in a large-scale V2G environment in which
EVs have the flexibility of getting service at multiple charging
stations coordinated by their individual aggregators, referred
to as MCS-MA in this paper, has not been explored so
far. Therefore, to fill this gap, we investigate the problem
of profit maximization under multiple CSs by proposing an
optimization model that incorporates the practically existing
constraints on the capacity and associated auxiliary costs of
each station. By considering the indeterministic arrival of EVs
during the scheduling process, we propose an online greedy
algorithm that guides incoming EVs to the most suitable CS.

We note that the most closing work to ours has been
addressed in [22] in which the authors investigate the problem
of profit maximization considering multiple geographically
distributed CSs. However, our work is different from [22] in the following ways: We consider multiple categories of PEVs
and investigate the maximization of relative obtainable profit
of both supply and demand entities. Furthermore, in contrast
to [22], more realistic system parameters are incorporated into
the proposed optimization model such as PEV’s battery as well
as the ancillary associated costs with CSs. Furthermore, we
achieve more insightful results through extensive simulations
under the problem objective such as the optimal number of
CSs under the proposed V2G system.


%

%

%

\ifCLASSOPTIONcaptionsoff
  \newpage
\fi

\end{document}